\documentstyle[aps,epsf]{revtex}
\begin{document}
\large
\baselineskip=20pt
\title{Relaxation and Diffusion for the Kicked Rotor.}
\author{Maxim Khodas and Shmuel Fishman}
\address{Physics Department, Technion, Haifa 32000, Israel}
\date{\today}
\maketitle
\begin{abstract}
The dynamics of the kicked-rotor, that is a paradigm
for a mixed system, where the motion in some parts of phase space is chaotic and in other parts is regular is studied statistically.
The evolution ( Frobenius-Perron ) operator of phase space densities
in the chaotic component is calculated in presence of noise,
and the limit of vanishing noise is taken is taken in the end of 
calculation.
The relaxation rates ( related to the Ruelle resonances )
to the invariant equilibrium density are calculated analytically
within an approximation that improves with increasing stochasticity.
The results are tested numerically.
The global picture of relaxation to the equilibrium density
in the chaotic component when the system is bounded and of diffusive
behavior when it is unbounded is presented. 
\end{abstract}
\ \ \ \ \ \ \ \ \ \ \ PACS number(s): 05.45.-a, 05.45.Ac

Statistical analysis is most appropriate for the exploration 
of global properties of systems that exhibit complicated 
dynamics ~\cite{Ott,Avez,Gaspardb,Dorfman}. For chaotic systems, 
$\hat{U}$,
the evolution operator of distributions of phase space trajectories
that is sometimes called the Frobenius-Perron (FP) operator
describes the statistical properties of the dynamics.
For many idealized systems exponential relaxation to the equilibrium density 
takes place.
This was established rigorously for hyperbolic systems (A systems),
like the baker map ~\cite{Avez,Gaspardb,Dorfman,Ruelle,HS}. 
The relaxation rates
related to the Ruelle resonances that are poles of the matrix elements
of the resolvent $\hat{R}=(z-\hat{U})^{-1}$ in a space of functions that are sufficiently smooth ~\cite{Ruelle}.
These poles are inside the unit circle in the complex $z$ plane while the 
spectrum of $\hat{U}$ is confined to the unit circle because of unitarity.
Most physically realistic models are not hyperbolic, but mixed where the phase space consists of chaotic and regular components.
For mixed systems sticking to regular regions takes place.
If the regular regions are small this effect is negligible for finite time,
that may be long, much longer than the time relevant to the experiment.
In this letter the FP operator and the relevant approximate relaxation rates are calculated analytically and numerically for the chaotic component
of the kicked rotor, that is a mixed system ~\cite{Thework}.

The kicked rotor is a paradigm ~\cite{1} for chaotic behavior of 
systems where one variable may be either bounded or unbounded in 
phase space.
If it is unbounded diffusion is found for the classical system 
~\cite{1,Balescu}.
In quantum mechanics this diffusion is suppressed by a 
mechanism similar to Anderson localization ~\cite{10}.
The kicked rotor is defined by the Hamiltonian
\begin{equation}
{\cal{H}}=\frac{1}{2}J^{2}+K cos\theta \sum_{n}\delta(t-n) , 
\label{0}
\end{equation}
where $J$ is the angular momentum, $\theta$ is the conjugate angle
$(0 \leq \theta < 2\pi)$ and $K$ is the stochasticity parameter.
Its equations 
of motion reduce to the standard map $\bar{\theta}=\theta+\bar{J}$ 
and $\bar{J}=J-K sin\theta$,
where $(\theta, J)$ and $(\bar{\theta}, \bar{J})$ are the angle 
and the angular 
momentum before a kick, and
just before the next 
kick respectively. For $K > K_{c} \approx 0.9716$ diffusion in phase space 
was found.

In the present paper the FP operator will be calculated for
the kicked rotor on the torus: \\ $(0\leq J<2\pi s~; 0 \leq \theta < 2\pi)$,
where $s$ is integer. 
The operator is defined in the space spanned by the 
Fourier basis:
\begin{equation}
\phi_{km}=(J\theta|km)=
\frac{1}{\sqrt{2\pi}}\frac{1}{\sqrt{2\pi s}}
\exp (im\theta)\exp \left( i\frac{kJ}{s} \right).
\label{3}
\end{equation}
The FP operator was studied rigorously for the hyperbolic systems 
and many of its
properties are known ~\cite{Avez,Gaspardb,Dorfman,Ruelle,HS}. It is a
unitary operator in ${\cal L}^{2}$,
the Hilbert space of square integrable functions. Therefore its resolvent
\begin{equation}
\hat{R}(z)=\frac{1}{z-\hat{U}}=
\frac{1}{z}\sum_{n=0}^{\infty}\hat{U}^{n}z^{-n} 
\label{5}
\end{equation}
is singular on the unit circle in the complex $z$ plane.
The matrix elements of $\hat{R}$ are discontinuous there and one finds a jump
between two Riemann sheets. 
The sum (\ref{5}) is convergent for $|z|>1$,
therefore it identifies the physical sheet, as the one connected with this
region. The
Ruelle resonances are the poles of the matrix elements of the resolvent, on
the Riemann sheet, extrapolated from $|z|>1$ ~\cite{HS}. These describe 
the decay of {\em smooth} probability distribution functions to the 
invariant density in a coarse grained form ~\cite{Gaspardb}. 
In spite of the solid 
mathematical
theory there are very few examples where the Ruelle resonances were calculated
 for specific systems ~\cite{Gaspardb,HS}. 
For the baker map it is easy to see that as  the resonances approach the unit 
circle, corresponding to slower decay, they are associated with coarser 
resolution in phase space ~\cite{HS}. 
For the kicked rotor (\ref{0}) the operation of $\hat{U}$ on a 
phase space density $\rho$ is
$\hat{U}\rho(\theta,J)=\rho(\theta-J,J+K sin(\theta-J))$.
To make the calculation well defined noise is added to the system. 
If noise that conserves J and leads to diffusion in $\theta$
is added to the free motion, 
the matrix elements of $\hat{U}$ in the basis (\ref{3}) are:
\begin{equation}
(k_{2}m_{2}|\hat{U}|k_{1}m_{1})=
J_{m_{2}-m_{1}}\left(\frac{k_{1}K}{s}\right)
\exp \left(-\frac{\sigma^{2}}{2}m_{2}^{2}\right)
\delta_{k_{2}-k_{1},m_{2}s}.  
\label{start}
\end{equation}
For $\sigma=0$ the operator is unitary as required.
It is 
shown explicitly that addition of the noise acts
effectively as coarse graining and the resulting 
evolution 
operator is not unitary 
(see also ~\cite{Shmuelinbook}). For large stochasticity
parameter $K$, it is shown here that in the Fourier basis (\ref{3}) 
the {\em slowest} 
relaxation modes, in the limit of infinitesimal noise, are found to be
identical to the 
modes of the diffusion operator ~\cite{Thework}.
Also the {\em fast} relaxation modes are calculated analytically 
in the present work, and the 
approximate analytical results are tested numerically ~\cite{Thework}.
These modes are not related to the spectrum of the 
FP operator that is
confined to the unit circle.
We believe we found all relaxation rates for distribution functions
that can be expanded in terms of the 
basis functions (\ref{3}). The 
immediate question is how is it possible that this description, that was 
established only for hyperbolic systems, holds for a mixed system. 
It is clearly 
approximate, and holds for large values of the stochasticity parameter 
$K$, since then most of the phase space is covered by the chaotic 
component. 
The physical reason for the decay of correlations
is, that in a chaotic system, the stretching and 
folding mechanisms lead to a persistent flow in the direction of 
functions with finer details, namely larger $|k|$ and $|m|$ in our case.
Consequently the projection on a given function, for example one of the 
basis functions (\ref{3}) in our case, decays ~\cite{Haake}. 
The crucial point is that this 
function should be sufficiently smooth.
This argument should hold 
as an approximation
also for the chaotic 
component of mixed systems. 
For smaller values of $K$ the weight of the regular regions 
increases. In such a situation, in the limit of increasing resolution the 
resonances related to the regular component 
are expected to move 
to the unit circle in the complex $z$ plane, 
corresponding to the quasi-periodic motion, while the resonances associated 
with the chaotic component stay inside the unit circle ~\cite{HW}. 

How is the FP operator related to the quantum mechanical evolution operator? 
It was shown numerically for the baker map
that if both operators are calculated with finite resolution they exhibit the 
same Ruelle
resonances ~\cite{Shmuelinbook}. 
Noise and coarse graining are introduced in field theoretical treatment of chaotic systems ~\cite{Zir,AAAS}.
Since the FP operator plays an important role in these theories
 our work is of relevance there. It also justifies some assumptions made in the calculation of the typical localization length for the kicked rotor
~\cite{Atland,Casati}.

We turn now to calculate the  
Ruelle resonances for the kicked rotor with the help of the
evolution operator (\ref{start}). 
The calculation will be done for finite noise $\sigma$ 
and then the limit  $\sigma \rightarrow 0$ will be taken.
These are the poles of matrix elements 
$R_{12}=(k_{1}m_{1}|\hat{R}(z)|k_{2}m_{2})$
of the resolvent operator $\hat{R}$ of (\ref{5})
when analytically 
continued from outside to inside the unit circle in the complex plane. 
It is useful to introduce 
$\hat{R}'(z)=1/(1-z\hat{U})=\sum_{n=0}^{\infty}z^{n}\hat{U}^{n}$ that 
is convergent inside the unit circle, because $||z\hat{U}||\leq 1$.
Its matrix elements are
$R^{'}_{12}=(k_{1}m_{1}|\hat{R}'(z)|k_{2}m_{2})=\sum_{n=0}^{\infty}a_{n}z^{n}$,
where
$a_{n}=(k_{1}m_{1}|\hat{U}^{n}|k_{2}m_{2})$.
The relation between the  matrix 
elements inside and outside of the unit circle implies that  
if $z_{c}$ is a singularity of $R_{12}$ then $1/z_{c}$ is a singular
point of $R^{'}_{12}$. Consequently
the first singularity of the analytic continuation of  $R^{'}_{12}(z)$ from 
inside to outside the unit circle gives the 
first singularity one incounters when analytically 
continuing  $R_{12}(z)$ from 
outside to inside the unit circle, i.e. it is just the leading
nontrivial resonance. 
It is determined 
from the fact that it is the radius of convergence $r$ of
the series for $R^{'}_{12}$ is given by
the Cauchy-Hadamard theorem: 
$r^{-1}=\lim_{n\rightarrow \infty}\sup\sqrt[n]{|a_{n}|}$
~\cite{CA}. 

The calculation of the coefficients $a_{n}=(k0|\hat{U}^{n}|k0)$ is performed
using the resolution of the identity
( introducing intermediate $|k_{i}m_{i})(k_{i}m_{i}|$ ),
and then substitution of (\ref{start}) and
summation over the $k_{i}$ leading to:
\begin{equation}
a_{n}=\sum_{m_{1}}\sum_{m_{2}}...
\sum_{m_{n-1}}
\prod_{l=1}^{n}
J_{M_{l}^{-}}\left(\frac{kK}{s}-KM_{l-1}^{+}\right)
e^{-(\sigma^{2}/2)m_{l}^{2}}
\delta_{M_{n-1}^{+},0} \ ,
\label{u14}
\end{equation} 
where $m_{0}=m_{n}=0$ while $M_{l}^{+}=\sum_{i=0}^{l}m_{i}$ and
$M_{l}^{-}=m_{l-1}-m_{l}$.
The calculation is performed for large $s$ and $K$ and
the limits are taken in order ~\cite{Thework}:
\begin{equation}
(1)\  s \rightarrow \infty,\ 
\label{u16}
(2)\ K \rightarrow \infty,\
(3)\  \sigma \rightarrow 0.
\end{equation}
For a sufficiently low mode so that $0<k K/s<<1$, the leading order term
in $K /s$ and $1/ \sqrt{K}$ is
\begin{equation}
a_{n}\sim 
\left[1-\frac{k^{2}K^{2}}{4s^{2}}\left(1-2J_{2}(K)e^{-\sigma^{2}}\right) 
\right]^{n}.
\label{u22}
\end{equation}
The resonance closest to the unit circle, $z_{k}$ is 
the inverse of the radius of convergence. Here
$z_{k}=e^{-(k^{2}/s^{2})D(K)}$, with
\begin{equation}
D(K)=\frac{K^{2}}{4}(1-2J_{2}(K)e^{-\sigma^{2}}),
\label{u25}
\end{equation}
that is just the value of the diffusion coefficient $D(K)$ found in ~\cite{RW}.
In the limit of $\sigma \rightarrow 0$
these are the relaxation rates in the diffusion equation.
The analysis of the off-diagonal matrix elements $a_{n}=(km|\hat{U}^{n}|k'm')$
leads to the same result.

In order to obtain the fast relaxation rates we have to 
calculate matrix elements that do not exhibit slow relaxation,
because such relaxation if present dominates the long time behavior.
For this purpose
we calculated the relaxation rates of disturbances from
invariant density that involve functions from
the subspace $\{ \ |0,m)\ \} $ with $m \neq 0$
and calculate $a_{n}=(0m|\hat{U}^{n}|km')$.
Again the resolution of the identity is introduced
( introducing intermediate $|k_{i}m_{i})(k_{i}m_{i}|$ ),
and summation over the $k_{i}$ yields a non vanishing result only if
$k/s \equiv q$ that is an integer. The expression for $a_{n}$
is found to be independent of $s$. 
The resulting resonances ( for large $K$ ) are
$\tilde{z}_{p}=\sqrt{ J_{2p}(p K)
\exp\left(-\sigma^{2}p^{2}/2\right) }$, 
where $p=|m|$ or $p=|\tilde{q}|$, where $\tilde{q}=q$ if $q \neq 0$
and $\tilde{q}=m'$ if $q=0$, depending which choice gives the larger 
absolute value ~\cite{Thework}. 
For $q=m'=0$, and $m \neq 0$ one 
finds $a_{n}=0$. If  $m=m'=q=0$ the only contribution is when all 
$m_{i}$ vanish and then $a_{n}=1$ for all $n$, resulting in the resonance
$z=1$, corresponding to equilibrium.

The FP operator is the evolution operator $\hat{U}$ in the limit
of vanishing noise. Therefore the Ruelle resonances are the poles of  
matrix elements of the
resolvent $ \hat{R} $ in this limit. They form several groups.
There is
$z_{0}=1$,
that is related to the equilibrium state. The resonances corresponding to the
relaxation modes related to the diffusion in the angular momentum are:
\begin{equation}
z_{k}=exp\left(-\frac{k^{2}K^{2}}{4s^{2}}(1-2J_{2}(K) )\right).
\label{resk}
\end{equation}
The resonances related to fast relaxation in the $\theta$ direction are:
\begin{equation}
\tilde{z}_{p}=\sqrt{J_{2p}(p K)}. 
\label{resp}
\end{equation}
In certain cases this result does not hold
for small intervals around $K^*$ ~\cite{Thework}.
The relaxation rates are 
$\gamma_k=|ln z_{k}|$ and $\tilde{\gamma}_p=|ln |\tilde{z}_{p}||$.

The analytical results that 
were obtained as the leading terms in an expansion in powers of $1/\sqrt{K}$
were tested numerically for finite $K$ and $\sigma=0$. 
For this purpose 
the correlation function
$C_{fg}(n)=(f|\hat{U}^{n}|g)$
was calculated numerically. 
For distributions $g$ and $f$ from the Fourier basis (\ref{3}), 
projected on the chaotic component,
the relaxation rates 
are expected to take the values $\gamma_k$ or $\tilde{\gamma}_p$.
For the diffusive modes one expects
$\gamma_k=(k^2/s^2)D(K)$,
where $D(K)$ is the diffusion coefficient (\ref{u25})
with $\sigma=0$. 
The values of $D(K)$ were
extracted from this relation
for various values of 
$k$ and $s$ and presented in Fig. 1.
\begin{figure}
\begin{center}
\begin{minipage}{10.1cm}
\centerline{\epsfxsize=8.0cm \epsfbox{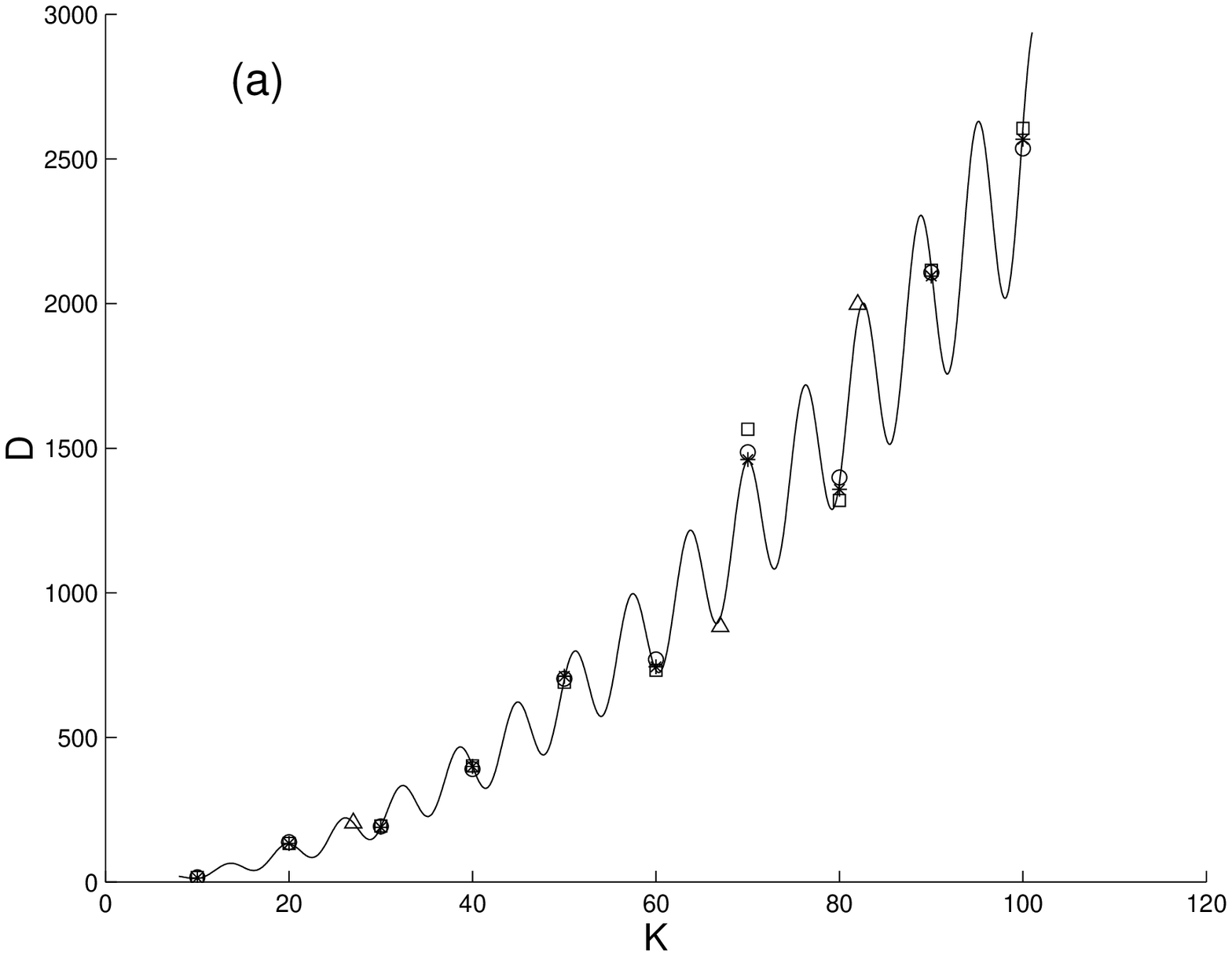}  }
\end{minipage}
\\
\begin{minipage}{10.1cm}
\centerline{\epsfxsize= 8.0cm \epsfbox{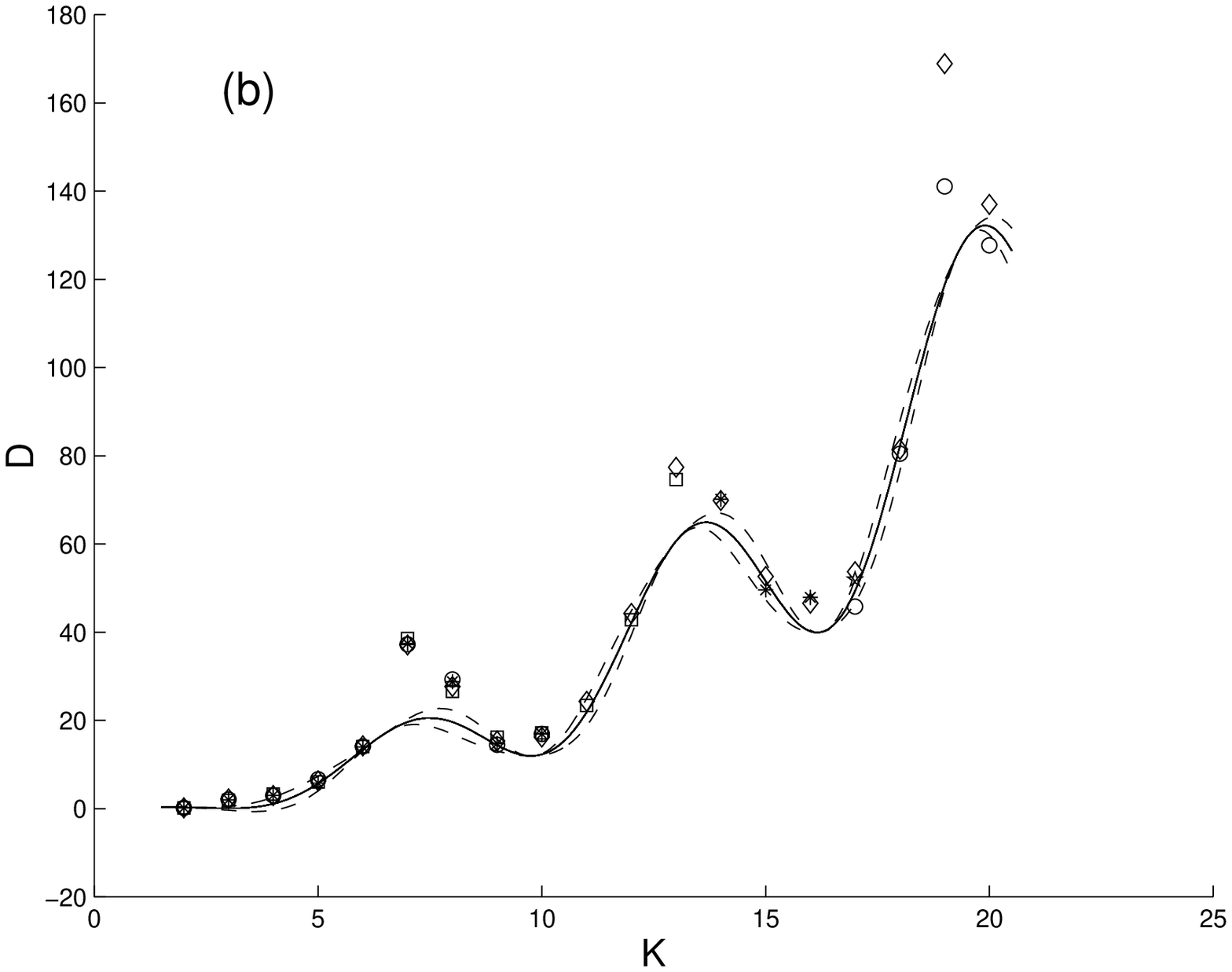}  }
\end{minipage}
\end{center}
\caption{ 
The diffusion coefficient $D$ as
extracted from the relaxation rates:
the first mode 
(squares) , the second mode  (stars), the fifth mode 
(circles) and off diagonal correlation functions
(triangles), compared to the
theoretical value (solid line).
The dashed line represents the approximate error. The values of $D$ obtained by
direct simulation of propagation of trajectories are marked by diamonds.}
\end{figure}
For large
values of $K$, excellent agreement with
the theory is found: the value of $D$ is found to be independent 
of $k$ and $s$ and
it agrees with (\ref{u25}).
For relatively smaller values of $K$,
the value of diffusion coefficient for some values of $K$ is larger than the one
that is theoretically predicted. 
The theoretical errors 
were estimated from the next term of the formula of Rechester 
and White for the diffusion coefficient ~\cite{RW}. 
In order to observe the rapidly relaxing modes the correlation
function $C_{fg}$ was calculated for
$g=\phi_{km'}$ so that $q=k/s$ is an integer and $f=\phi_{0m}$.
In Fig. 2 the numerical estimate for $\tilde{\gamma}_{p}$ is compared with 
the theoretical prediction obtained from (\ref{resp}). 
\begin{figure}
\centerline{\epsfysize 6.0cm \epsfbox{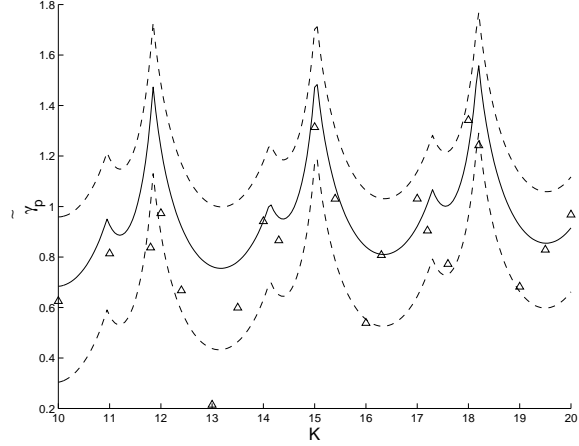}  }
\caption{ 
The fast relaxation rates $\tilde{\gamma}_p$ for 
$f=\phi_{01},g=\phi_{02}$ (triangles),
compared to the theoretical value (solid line).
The dashed lines denote the theoretically estimated error.
Here we used $ s=1 $ and $ N=10^{8} $. 
} 
\label{fig:fastmodes}	
\end{figure}
The error in the theoretical prediction is estimated as the value of the next 
order contribution to $a_{n}$.  
The main reason for disagreement between the theory and the numerical simulations is sticking to the islands of stability
and accelerator modes ~\cite{Zaslav}.

Finite noise leads to 
the effective truncation of the evolution operator (\ref{start}) . In the basis (\ref{3}) it means that it results in limited resolution. Moreover for $\sigma > 0$ the operator
$ \hat{U} $ is non unitary. 
The approximate eigenvalues of $ \hat{U} $ given by (\ref{start}) that were found
in this work are $1$, $z_{k}$ and 
$ \tilde{z}_{p}$. Because of the effective truncation, 
$ \psi_{\gamma} $, the eigenfunction of $ \hat{U} $, can be expanded in terms of the basis states (\ref{3}). The relaxation rates of these 
eigenstates are
$-ln(z_{k})$ and $ -ln(|\tilde{z}_{p}|)$. 
In the limit $ \sigma \rightarrow 0 $ the evolution operator is unitary, and
$ \psi_{\gamma} $ approach some generalized functions while   
$ z_{k} $ and $\tilde{z}_{p} $ approach the values (\ref{resk},\ref{resp}).
These are the Ruelle resonances similar to the ones found for hyperbolic 
systems such as the baker map ~\cite{HS}. 
Here noise was used in order to make the analytical calculations
possible. In real experiments some level of noise is present,
therefore the results in presence of noise are of experimental
relevance.

In summary, the Ruelle resonances, that were found rigorously for
hyperbolic systems can be used for an approximate description of 
relaxation and transport in the chaotic component of mixed systems. 
The relaxation of distributions in phase 
space 
to the invariant density
takes place in stages. First the inhomogeneity in $\theta$ 
decays with the rapid relaxation rates $\tilde{\gamma}_{p}$ and then 
relaxation of the inhomogeneities in the $J$ direction takes place 
with the relaxation rates of the diffusion equation. In the limit
$s\rightarrow \infty $ the inhomogeneity in $\theta $ relaxes and then 
diffusion in the momentum direction takes place. 

We have benefited from discussions with O. Agam, E. Berg, 
R. Dorfman, I. Guarneri, 
F. Haake, 
E. Ott, R. Prange, S. Rahav,  J. Weber and M. Zirenbauer. 
We thank in particular 
D. Alonso for extremely
illuminating remarks and helpful suggestions. This research was supported in part by the
US--NSF grant NSF DMR 962 4559, the
U.S.--Israel Binational Science Foundation (BSF), by the Minerva
Center for Non-linear Physics of Complex Systems, by the Israel
Science Foundation, and by the Fund for Promotion of Research at the
Technion. One of us (SF) would like to thank R.E. Prange for the
hospitality at the University of Maryland where this work was
completed.

\typeout{References}

\end{document}